\documentclass{ws-p8-50x6-00}

\begin{document}

\def\PL{{ \it Phys. Lett.} }
\def\PRL{{\it Phys. Rev. Lett.} }
\def\NP{{\it Nucl. Phys.} }
\def\PR{{\it Phys. Rev.} }
\def\MPL{{\it Mod. Phys. Lett.} }
\def\IJMP{{\it Int. J. Mod .Phys.} }

\title{Diluting Gravity with Compact Hyperboloids}

\author{Mark Trodden}

\address{Physics Department, Syracuse University, Syracuse NY 13244-1130,
USA\\E-mail: trodden@phy.syr.edu}  


\maketitle

\abstracts{I give a brief informal introduction to the idea and tests of large extra dimensions,
focusing on the case in which the space-time manifold has a direct product structure. I then 
describe some attractive implementations in which the internal space comprises a compact hyperbolic
manifold. This construction yields an exponential hierarchy between the usual Planck scale and 
the true fundamental scale of physics by tuning only ${\cal O}(1)$ coefficients, since the 
linear size of the internal space remains small. In addition,
this allows an early universe cosmology with normal evolution up to
substantial temperatures, and completely evades astrophysical constraints.
}
\vspace{-4mm}
\noindent
SU-GP-00/10-1

\section{Some Background about Large Extra Dimensions}
There are a number of talks in this conference about the idea of large extra dimensions. With this 
in mind I will begin with a brief overview of the general idea and of the constraints on it.

\subsection{Motivations : the Hierarchy Problem}
The hierarchy problem, for our purposes, is usually posed in the following way: Why is gravity so
much weaker than the other forces? To make this concrete compare
\begin{equation}
G_N=M_{\rm pl}^{-2} \sim 10^{-33} \ \ \mbox{GeV}^{-2}
\end{equation}
with
\begin{equation}
G_F=M_{\rm W}^{-2} \sim 10^{-5} \ \ \mbox{GeV}^{-2}
\end{equation}
Phrased in this way, the Planck mass $M_{\rm pl}$ is considered to be the fundamental scale of
physics and the puzzle is the comparative smallness of $M_{\rm W}$, or the comparative strength of
the electroweak force. The real problem here is not so much the fact that the two scales differ
so drastically, but rather that such an arrangement is doomed to be ruined by radiative corrections
in a generic quantum field theory. Thus, theoretical efforts have mostly focused on restoring the
{\it technical} naturalness of this vast difference is scales via, for example, supersymmetry, in
which the offending quadratic divergences are absent.

Given this suggestive phrasing, a logical alternative statement of the problem becomes clear. If we
instead imagine that the fundamental scale of physics $M_*$ is close to the weak scale $M_{\rm W}$, 
then the hierarchy problem becomes: Why is $M_{\rm pl}$ so much larger? When considering this 
possibility however, one must keep in mind some important facts about physics at these scales:

\begin{itemize}
\item The Electroweak interactions have been tested up to energies \( E\sim M_{W} \), or equivalently, 
down to scales
\( d\sim M_{W}^{-1}=10^{-16}\textrm{cm} \)
\item Gravity, in the form of Newton's law has been tested down to a scale 
\( d\sim 10^{33}/M_{\rm{pl}}\sim 1\textrm{cm} \) 
\end{itemize}
Thus, we know that there are no quantum gravity effects up to energies \( E\sim M_{W} \), since,
for example, there is no evidence of energy lost to gravitons in \( e^{+}e^{-} \)
annihilations.

Given these constraints, let us examine how it might be that the Planck mass is a comparatively 
huge derived quantity in a theory in which the weak scale is fundamental. The fundamental idea 
behind the new constructions is the following\cite{Anton,ADD}. 
Imagine that space time has $3+1+d$ dimensions, and 
that $d$ of these are compact while the remaining $3+1$ comprise the familiar space-time in which
we appear to live. Now make the following two crucial assumptions:
\begin{itemize}
\item Standard model particles are {\emph{confined}} to the \( 3+1 \) dimensional sub-manifold.
\item Gravity is not confined, and therefore gravitons propagate in the bulk
\end{itemize}
These assumptions will turn out to be crucial, since to obtain the necessary hierarchy we will require 
large volume extra dimensions, and these are ruled out by the precision tests mentioned above if
the weak interactions can take place in the bulk. While such a structure may seem somewhat unnatural
from the phenomenological viewpoint taken here, it is important to mention that the appropriate 
behavior occurs readily in some compactifications of string theory. In particular, in 
Ho{\v r}ava-Witten theory\cite{HW}, precisely this 
division of interactions occurs. The essential M-theoretic ingredient responsible for this is the 
existence of D-branes, non-perturbative extended objects, in the string spectrum\cite{polchinski}. 
Open strings must
end on D-branes (with ``D''irichlet boundary conditions on their endpoints), while the open strings 
are free to propagate through the whole larger space. Since the excitations of open strings 
correspond to standard model-like degrees of freedom, while the closed string excitations describe
the geometric (gravitational) degrees of freedom, the resulting structure is one with standard model
fields confined to a D-brane, and gravity propagating in the full space.

\begin{figure}[h]
\epsfxsize=20pc 
\center \epsfbox{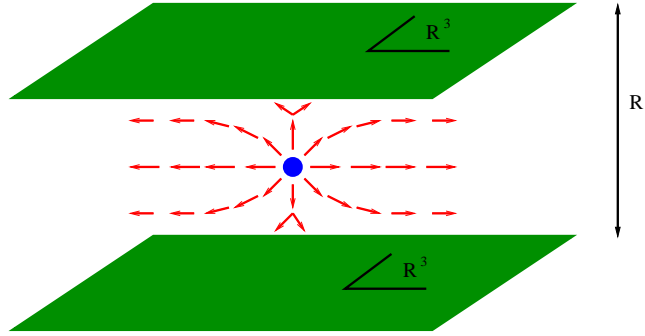} 
\caption{Gravitational flux around a point mass in a direct product space with one extra dimension.
\label{flux}}
\end{figure}

To be concrete, let us describe the original large extra dimension scenario. Imagine that the full
space-time manifold is
\begin{equation}
{\cal M}^{3+1} \times T^d \ ,
\end{equation}
where ${\cal M}^{3+1}$ describes a flat Minkowski or cosmological space which we observe, and $T^d$
is a d-dimensional torus of common linear size $R$ describing the internal space. Consider how 
gravitational flux spreads
out around a point mass $m$, as in figure~\ref{flux}. 

The gravitational
acceleration experienced by a test particle at distance $r$ from such a point mass obeys 
the following
\begin{eqnarray}
|{\bf g}| & = & \frac{G_{(4+d)}m}{r^{2+d}} \ \ \ \ \ \mbox{for $\ \ \ r \ll R$} \nonumber \\
& = & \frac{G_{(4+d)}m}{R^d r^{2}} \ \ \ \ \ \mbox{for $\ \ \ r \gg R$} \ .
\end{eqnarray}
This is easy to understand in terms of Gauss' law : On small scales $r\ll R$ gravity is diluted by 
spreading out into all the dimensions, whereas on large scales $r\gg R$ gravity is smeared out over
the internal space and can only spread into the $3+1$ dimensional space. 

Now, since we understand gravity extremely well on large scales, we must identify
\begin{equation}
G_{4} \equiv \frac{G_{(4+d)}}{R^d} \ \ \ \ \Leftrightarrow \ \ \ \ 
M_{\rm pl}^2 \equiv M_*^2 R^d \ .
\label{flatMpl}
\end{equation}
Thus, the observed Planck mass can be huge, even if the fundamental scale of physics is of order
a TeV. All that is needed is that the {\it volume} of the extra dimensional space (here $R^d$) is
large enough.  In particular, for $M_* \sim 1$ TeV, we require
\begin{equation}
R\sim 10^{\frac{32}{d}-17} {\rm cm} \ .
\end{equation}
For example, 

\begin{itemize}
\item If $d=1$, we need $R\sim 10^{13}$ cm, which is obviously excluded by any number of large scale 
tests of gravity.
\item If $d=2$, we need $R\sim$ mm, a truly large extra dimension.
\end{itemize}
So, in a picture with extra dimensions and standard model particles restricted to the brane, the 
hierarchy problem can be recast in the interesting guise of a mismatch of spatial scales, rather than 
one of energies.

\subsection{Constraints}
\begin{figure}[h]
\epsfxsize=20pc 
\center \epsfbox{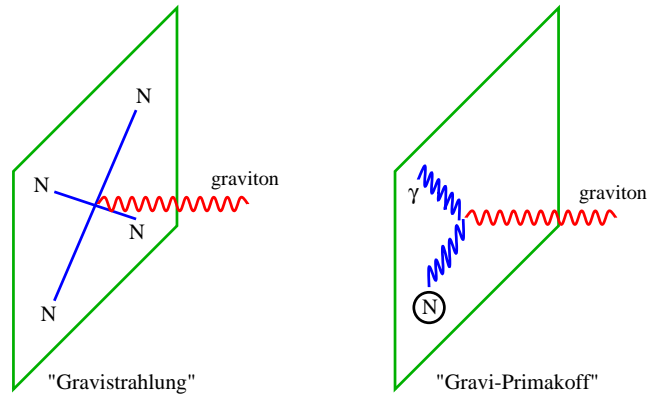} 
\caption{Two processes through which astrophysical bodies can lose energy to KK gravitons.
\label{astroconstraints}}
\end{figure}

The constraints on such a possible structure for space-time and particle physics come from three main
sources: the laboratory, astrophysics and cosmology. In the laboratory\footnote{For a nice summary
of laboratory constraints see \cite{Landsberg}}, the relevant quantity is 
the amplitude for single graviton emission:
\begin{equation}
{\cal A}\sim \sqrt{M_*^{-(d+2)}} \ .
\end{equation}
This allows one to write the dimensionless graviton emission rate for a process of energy $\Delta E$
as
\begin{equation}
{\cal R} \sim G_{(4+d)} \Delta E^{d+2} \equiv \left(\frac{\Delta E}{M_*}\right)^{d+2} \ .
\end{equation}
There are two important facts that we shall need from this expression. One is that the constraint 
becomes weaker the more extra dimensions there are, and the second is that the constraint is stronger
for higher values of $\Delta E$. As it turns out, collider constraints \cite{collider} 
are currently not the strongest
ones faced by the large extra dimension picture. This can be seen, for example, by considering the
process $K \rightarrow \pi + $ graviton,
\begin{figure}[h]
\epsfxsize=15pc 
\center \epsfbox{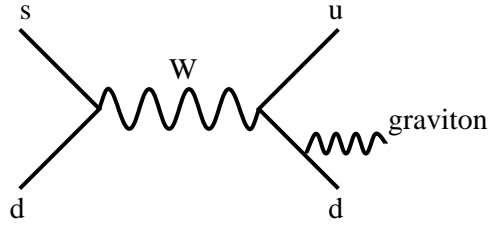} 
\caption{A feynman diagram for $K \rightarrow \pi + $ graviton.
\label{feynman}}
\end{figure}
for which the branching ratio is
\begin{equation}
{\rm Br}(K\rightarrow \pi g) \sim \left(\frac{m_K}{M_*}\right)^{d+2} \sim 10^{-12} \ \ \ \ \ 
\mbox{for \ \ \ $d=2$} \ .
\end{equation}
More important are constraints arising from astrophysics. The basic issue is that the Kaluza-Klein (KK)
modes in the model are light $M_{KK} \geq R^{-1}\geq 10^{-4}$eV, and numerous $N_{KK}\simeq 
M_{\rm pl}^2/M_*^2 \leq 10^{32}$. Thus, although each KK mode is very weakly coupled, of order 
$1/M_{\rm pl}$, to standard model particles on the brane, there are so many of them that they can 
be copiously produced by energetic processes on the brane. Therefore, it is possible for 
astrophysical bodies to lose energy by emitting gravitons into the extra dimensions. The most
important
ways in which this can occur have been termed {\it gravistrahlung}, referring to the production
of a KK graviton in heavy nucleon collisions, and the {\it gravi-Primakoff} process, in which
a KK graviton is produced through a photon scattering from a heavy nucleus 
(see figure~\ref{astroconstraints}).
As mentioned earlier, the most energetic processes yield the tightest constraints, and so it is 
useful to consider energy loss from the most energetic astrophysical event yet known, supernova 1987A
(SN1987A). A careful analysis\cite{KKconstraints} yields
\begin{equation}
M_* \geq 50 \ \ \mbox{TeV}\ \ \  \mbox{for $d=2$}
\end{equation}
(and less important constraints on higher values of $d$).
As is often the case in modern particle physics, some of the most important constraints on this 
structure arise from cosmological considerations. I will just mention two main issues here. As I have
described, extra dimensions provide an alternative mechanism through which astrophysical bodies can 
cool. This remains true for the universe itself. In the standard cosmology, the universe cools
adiabatically as the scale factor $a(t)$ increases, with the temperature-time relationship (in the 
radiation-dominated era)
\begin{equation}
\frac{{\dot T}}{T}\sim \frac{{\dot a}}{a}=H\sim \frac{T^2}{M_*} \ .
\end{equation}
In the new picture, there is a competing evaporative cooling mechanism due to 
{\it cosmic gravistrahlung}, with temperature-time relationship
\begin{equation}
\frac{{\dot T}}{T}\sim \frac{T^{d+3}}{M_*^{d+2}} \ .
\end{equation}
Since the standard cosmology is so successful as a description of the universe at temperatures below 
that of, for example, primordial nucleosynthesis, we require that the new cooling mechanism be
sub-dominant at least at temperatures below that. To this end it has become customary to define the
{\it normalcy temperature} $T_*$, to be that temperature below which evaporative cooling is 
negligible compared to the usual adiabatic mechanism. In the basic large extra dimension scenario that 
I am reviewing here, this temperature is
\begin{eqnarray}
T_* & \sim & 10 \ \mbox{MeV}\ \ \ \mbox{for $d=2$} \nonumber \\
& \sim & 10 \ \ \mbox{GeV}\ \ \ \mbox{for $d=6$ (string theory)} \ .
\end{eqnarray}
The second cosmological constraint arises from the possibility of gravitino overproduction. If the 
underlying theory is supersymmetric, one expects gravitinos to be thermally produced. If they are
produced at temperature $T$, they have a lifetime
\begin{equation}
t\sim \frac{M_{\rm pl}^2}{T^3} \ .
\end{equation}
If $M_*$ is too small, these particles may over-close the universe or distort the $\gamma$-ray
background. These constraints are most relevant for $d=2$, in which case they yield
\begin{eqnarray}
M_* & > & {\cal O}({\rm TeV}) \nonumber \\
& > & 110 \ \ {\rm TeV} \ .
\end{eqnarray}

To summarize the combined constraints from the laboratory, astrophysics and cosmology; $d=1$ is 
experimentally ruled out (easily), $d=2$ is quite constrained, and $d>2$ is relatively unconstrained.

\section{Some Interesting Manifolds}
I have spent some time giving an overview of the basic concepts behind and constraints on the large extra
dimension model, as originally proposed. I would now like to shift gears a little, and describe how
these ideas might be extended to the case when the internal manifold is no longer flat 
(merely a torus)\cite{our paper}.
More specifically I will be interested in topologically nontrivial internal spaces, and in particular in
the case that the internal space is a compact hyperbolic manifold.

Compact hyperbolic manifolds (CHMs)\cite{Thurston}, are obtained from $ H^{d}$, the universal covering 
space of hyperbolic 
geometry (that admitting constant negative curvature), by modding out by an appropriate freely acting discrete
subgroup $ \Gamma  $ of the isometry group of $H^{d}$. (If $\Gamma$ is not freely-acting,
then the resulting quotient is a non-flat non-smooth orbifold. Such a structure may be related to the 
Randall-Sundrum models\cite{RS})  

Consider space-times of the form 
\begin{equation}
{\cal M}^4\times (H^d/\Gamma|_{\rm free}) \ ,
\end{equation} 
with ${\cal M}^4$ a Friedmann, Robertson-Walker (FRW) 4-manifold, with metric
\begin{equation}
G_{IJ}dz^I d z^J = g^{(4)}_{\mu\nu}(x)dx^{\mu}dx^\nu + R_c^2 g^{(d)}_{ij}(y)
dy^{i}dy^j.
\label{metric}
\end{equation}
In this expression $R_c$ is the physical curvature radius of the CHM, 
so that $g_{ij}(y)$ is the metric on the CHM normalized so that
its Ricci scalar is ${\cal R}=-1$, and $\mu = 0,\ldots,3$, $i= 1,\ldots,d$. 

Locally negatively curved spaces exist only for $d \ge 2$, and the properties of CHMs are well 
understood only for $d\leq3$. However, it is known that CHMs
in dimensions $d\geq 3$ possess the important property of
{\em rigidity} \cite{MostowPrasad}.
As a result, these manifolds have {\em no massless shape moduli}, implying that  
the volume of the manifold, in units of the curvature radius $R_c$,
cannot be changed while maintaining the homogeneity of the geometry.
Therefore, the stabilization of such internal
spaces reduces to the problem of stabilizing a single modulus,
the curvature length or the ``radion''.

\begin{figure}[h]
\epsfxsize=25pc 
\center \epsfbox{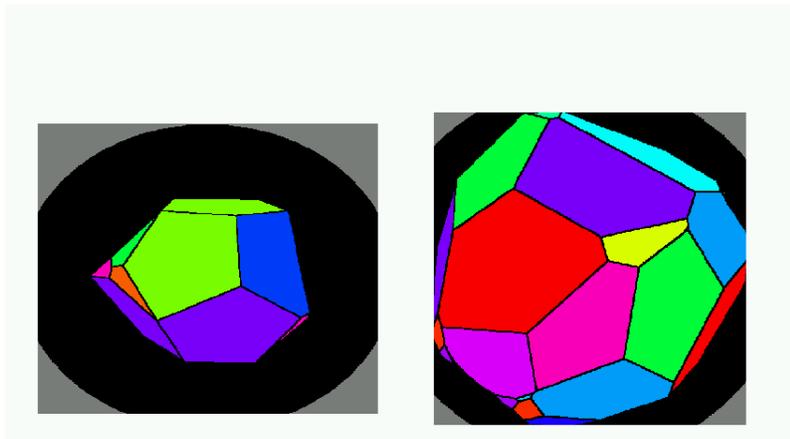} 
\caption{Two CHMs, one small volume, the other large volume.
\label{CHMs}}
\end{figure}

Although CHMs may seem like quite abstract objects, their popularity among mathematicians has
led to some useful tools for visualizing their structure. In figure~\ref{CHMs} I have used the
Geomview package\cite{geomview}to display two examples of CHMs generated by Jeff Weeks' SnapPea 
program\cite{snappea}. 
These are presented
in the Poincar{\' e} metric, and what is important here is that those parts of the manifold that are
identified under actions under $\Gamma$ are shaded in the same way. I would just like to draw 
attention to two important features. First, the second example has a much larger volume than the first in 
units of the curvature radius. In addition, the second example is considerably more topologically complex than 
the first, as can be seen by the much higher number of identifications under $\Gamma$. These features are 
interrelated, and are quite general.

\subsection{Compact Hyperbolic Spaces and Volume}
The central feature of CHMs that I will exploit here is the behavior of their volume as a function
of linear size in the manifold. 
A specific example that it  is useful to keep in mind is a 3-sphere
of radius $r$, cut out of an $H^3$ of curvature radius $R_c$.  
The volume of such a sphere can be calculated exactly, and is given by
\begin{equation}
\label{threevolume}
{\rm Vol}(r)= \pi R_{c}^{3}\left[\sinh \left(\frac{2r}{R_c}\right)-\frac{2r}{R_c}\right] \ .
\end{equation}
It is useful to examine this expression in two limits. When $r\ll R_c$, the first term in a Taylor 
expansion yields 
\begin{equation}
{\rm Vol}(r)\sim \frac{4}{3}\pi r^3 \ ,
\end{equation}
as one would expect, since the manifold looks flat on these scales. However, in the opposite limit 
$r\gg R_c$, we obtain
\begin{equation}
{\rm Vol}(r) \sim \frac{\pi}{2} R_c^3 e^{2r/R_c} \ ,
\end{equation}
and therefore the volume of the sphere grows exponentially with linear size for large radius.

This result remains true for other compact spaces constructed from $H^3$.
In general, the total volume of a smooth compact hyperbolic
space in any number of dimensions is
\begin{equation}
{\rm Vol(CHM)}=R_c^d~ e^{\alpha } \ ,
\end{equation}
where $\alpha$ is a constant, determined by topology.
Since the topological invariant $e^{\alpha}$ characterizes
the volume of the CHM,
it is also a measure of the largest distance $L$ around the manifold.
Although CHMs are globally anisotropic,
since the largest linear dimension gives the most significant contribution to
the volume, one can employ eq.~(\ref{threevolume}), or its generalizations
to $d\neq3$, to find an approximate relationship between
$L$ and ${\rm Vol(CHM)}$.
For $L \gg R_c/2$ the appropriate asymptotic relation,
dropping irrelevant angular factors, is
\begin{equation}
e^\alpha \simeq \exp\left[\frac{(d-1) L}{R_c} \right] \ .
\end{equation}
In a model with such a manifold as an extra dimensional space, this relationship allows us to 
compute the expression for $M_{\rm pl}$ in terms of linear distance in the space, to compare with
(\ref{flatMpl}) in the flat case. The relevant expression is
\begin{equation}
M_{\rm pl}^2=M_*^{2+d}R_c^d e^{\alpha } \simeq M_*^{2+d} R^d_c \exp\left[\frac{(d-1) L}{R_c} \right] \ .
\end{equation}
Thus, in strong contrast to the flat case, in which $M_{\rm pl}$ has a power law dependence on
linear size, with a CHM the relationship is exponential.
As I'll mention later, the most reasonable and interesting case is the smallest possible curvature radius,
$R_c \sim M^{-1}_*$, since this is the only scale available in the problem.  
Taking $ M_*\sim $~TeV then yields
\begin{equation}
\label{lmaxnum}
L\simeq 35 M_*^{-1} = 10^{-15} {\rm mm} \, .
\end{equation}
Therefore, one of the most attractive features of a CHM internal space 
is that {\it to generate an exponential
hierarchy between $M_*\sim$~TeV, and $M_{\rm pl}$ requires only that the linear size
$L$ be very mildly tuned}.

\subsection{Eigenmodes and Kaluza-Klein Excitations}
I have tried to convince you that CHMs provide an attractive alternative manifold for implementing
large extra dimension ideas. However, if this idea is to be taken seriously it is necessary to examine
the constraints and possible experimental signatures. 

To uncover the relevant physics of these models one must consider the spectrum
of small fluctuations $h$ in the metric around the background metric. Writing
\begin{equation}
G_{IJ} \to G_{IJ} + {\rm e}^{ip.x} h_{IJ}(y) \,.
\end{equation}  
one sees 3 different types of KK fluctuations 
\begin{itemize}
\item $h_{\mu\nu}$, the spin-2 piece;
\item $h_{ij}$, with indices only in the internal directions, giving
spin-0 fields for the 4D observer;  
\item $h_{i\mu}$, the mixed case, giving spin-1 4D fields.  
\end{itemize}
The 4D KK masses of these states are the eigenvalues of the appropriate internal-space Laplacians 
acting on $h(y)$.
For the spin-2
case the relevant operator is the Laplace-Beltrami operator $\Delta_{LB}$
(the Laplacian on scalar functions in the internal space),
defined by
\begin{equation}
\Delta_{LB} \phi(y) = |g(y)|^{-1/2} \partial_i \left(
|g(y)|^{1/2} g^{ij}  \partial_j\phi(y)  \right) .
\label{beltrami}
\end{equation}

Although there are no known analytic expressions for the individual
eigenvalues of $\Delta_{LB}$ on a CHM of any dimension, some generic properties are known.
First, a variational argument shows that the spectrum of
$\Delta_{LB}$ is bounded from below,
and the lowest eigenmode is just the constant function on the CHM.
This zero mode is the internal space wave-function
of the massless spin-2 4D graviton.  

Second, since the internal space in compact, the
spectrum of $\Delta_{LB}$ on a CHM is discrete and has a gap between
the zero mode and the first excited KK state.
The size of this gap is all important. A crucial point is that
most of the eigenmodes of $\Delta_{LB}$ on a
CHM have wavelengths less than
$R_c$, and the number density of these modes is well approximated
by the usual Weyl asymptotic formula\footnote{There can also be a few 
lighter {\em supercurvature modes},
with wavelengths as large as the longest
linear distance in the manifold, and masses thus
bounded below by $L^{-1}$.}
\be
n(k)= (2\pi)^{-d} \Omega_{(d-1)} V_d k^{d-1} \ ,
\label{weyl}
\ee
where $\Omega_{(d-1)}={\rm Area}(S^{d-1})$.
Further, bounds on the value of the first non-zero eigenvalue
are known.  In the best-studied CHM case of $d=2$ it can be proven that
a large enough volume (and thus genus) $d=2$ CHM
will have first eigenvalue $\ge 171/(784R_c^2)$.  
The analogous conjecture 
in $d=3$ is more problematic, but has also been made~\cite{Brooks}.
In addition, numerical studies of the spectra of even small volume $d=3$ CHMs
show that they have very few modes with $\lambda < R_c$ \cite{numeric}.

The basic result is that the first KK modes are
exponentially more massive than those in the flat case.  
This drastically raises the threshold for their production, and as a consequence
the astrophysical
bounds\cite{ADD3,KKconstraints} completely disappear since the lightest KK mode
has a mass (at least 30~GeV),
much greater than the temperature of even the hottest astrophysical object.
Similarly the large KK masses imply a much higher normalcy temperature $T_*$,
up to which the evolution of our brane-localized 4D universe can
be normal radiation-dominated FRW.

Turning to the spin-0(1) excitations, the detailed form of the Laplacian
is modified. 
However, the Mostow-Prasad rigidity theorem for CHMs of dimension
$d\ge 3$ implies that $\Delta_{\rm LL}$ has no zero modes, and it is
conjectured that the gap to the first excited state is of similar size to
the Laplace-Beltrami case, a result
that is physically reasonable.  Finally for the spin-1 fluctuations $h_{i\mu}$
recall that any zero modes would correspond to KK gauge-bosons
 and are directly related to the continuous isometries of the compact space.
But, as a result of the quotient by $\Gamma$, CHMs have
no such isometries, and thus there are no massless KK gauge
bosons.  The non-zero
KK modes once again have a mass gap that is at least as large as $1/L$ and is more
likely close to $\sim 1/R_c$, as in the previous cases.
Thus these additional types of fluctuation should not disturb our results.

\subsection{Radion Stabilization}
In order for the CHM model to work, it is necessary to realize $R_c \sim M_*^{-1}$ and
$e^\alpha \simeq \exp\left((d-1) L/R_c\right)\gg 1$ consistently with the ansatz of a 
factorizable geometry, a static internal space, and the vanishing of the
4D long-distance ($\gg L$) cosmological constant (CC) $\Lambda_4$. To see how this might work,
consider a 3-brane embedded in $(4+d)$ dimensions, with bulk and brane actions
\begin{eqnarray}
& S_{\rm bulk}  &= \int d^{4+d}x
\sqrt{-|g_{(4+d)}|} \biggl( M_*^{d+2}{\cal R} +\Lambda
- {\cal L}_m \biggr) \\
& S_{\rm brane}  &=  \int d^{4}x \sqrt{-|g_{(4)}^{\rm induced}|}
\biggl( f^{4} + \ldots \biggr),
\label{action}
\end{eqnarray}
respectively, where ${\cal L}_m$ is the bulk matter field Lagrangian, and $f^4$ is the brane 
tension. Note that, since CHMs are just quotients of $H^d$, there will exist
a uniform negative bulk cosmological constant $\Lambda \sim M_*^{4+d}$, and that
to ensure a static internal space, this must be balanced in the field equations by
the small curvature radius of the internal space.
Dimensionally reducing these actions yields an effective 
4D potential for the radion $R_c$ of the form
\begin{equation}
V(R_c) = \Lambda R_c^d e^\alpha - M_*^4 e^\alpha (M_* R_c)^{d-2}
+W(R_c) +f^4\ .
\label{potential}
\end{equation}
Here the first two terms arise from the $(4+d)$ bulk CC term,
and the curvature of the internal space.  Now, in general we may
expand $W(R_c)$, which comes from ${\cal L}_m$,
as a Laurent series in $R_c$
\be
W(R_c)=\sum_{p} a_p \frac{M_*^4}{(R_c M_*)^p} \ ,
\ee
with dimensionless coefficients $a_p$. Broadly speaking, there are then two
interesting possibilities.
If the determination of the minimum is dominated by a competition
between any {\em two} terms in $V$, then the condition
that the 4D CC vanish ($V_{\rm min}=0$) cannot be achieved with a brane tension
such that $|f^4| \le M_*^4$. Thus, such a situation is not consistent with
the basic assumption that a low-energy effective
theory is valid on the brane

Fortunately there is an attractive alternative.  If {\em three} or more
$R_c$-dependent terms in $V(R_c)$ are all important at the minimum
(for example the CC and curvature terms, and one of the matter
terms from $W$) then we can tune the coefficients $a_p$ such that
$V_{\rm min}=0$, without needing $f^4\gg M_*^4$.  Thus, the basic
assumptions remain consistent.  Moreover, this tuning is
particularly natural in the CHM case precisely because the
minimum will naturally occur for a curvature radius close to the fundamental
scale $R_c \sim M_*^{-1}$, at which the high-scale theory will
produce many different terms that contribute roughly in an equal way.
(This is exactly the opposite situation from the large flat extra dimension
case where the minimum has to occur at a length scale much greater than
$M_*^{-1}$.) This one fine-tuning is just the usual
4d CC problem. It seems unlikely to me that this one fine tuning will be solved
within these models, since in the end one is left with an arbitrary higher-dimensional 
cosmological constant that one can add to the theory. Nevertheless, at the very least 
the cosmological constant problem appears in a different guise in these models.

It remains to check one important detail.
In the usual large extra dimension scenario the radion moduli problem in the early universe
provides quite a strong constraint\cite{ADKMR}.  In the CHM case this problem is much weakened.
The radion, which is the light mode
corresponding to dilations of the internal space, gets its mass from
the stabilizing potential $V(R_c)$. Here
\begin{equation}
m_r^2 = \frac{1}{2} \frac{R_c^2 V''(R_c)}{e^\alpha M_*^{d+2} R_c^d}
\simeq \frac{1}{R_c^2} \ ,
\end{equation}
which is close to $M_*^2\sim{\rm TeV}^2$.
Therefore, the radion lifetime is
$t \sim M_{\rm pl}^2/M_*^3$, 
much shorter than in the case of flat extra dimensions, 
and only slightly longer than the age of the universe at nucleosynthesis,
even if $M_*\sim {\rm TeV}$.

\section{Conclusions and Further Directions}
I have briefly reviewed the structure of theories with large extra dimensions, and
discussed the main constraints on these models. I have then described an important
modification to these theories, in which the internal manifold comprises a CHM. With
this modification, the hierarchy problem is solved by a mild tuning of parameters, in 
an interesting and topologically stable way.
 
While cosmologically and astrophysically much safer,
models with internal compact hyperbolic spaces
do have testable signatures accessible to
collider experiments.  Since KK modes 
abound close to the fundamental scale, Standard Model
particle collisions with center-of-mass energies near this scale
will result in the production of KK particles,
detectable by a distinctive missing  energy signature \cite{collider}.
Although this is qualitatively similar to the scenario of \cite{RS},
the spectrum of KK modes one will see is quite distinctive.

A full exploration of these experimental signatures
will require a more complete investigation of the spectrum of large CHMs,
in particular the issues of isospectrality
and homophonicity of such manifolds.
It is quite likely that such CHMs have other
implications for higher-dimensional physics.

\section*{Acknowledgements}
I would first like to acknowledge my collaborators on CHMs: John March-Russell, Nemanja Kaloper 
and Glenn Starkman.
I would also like to thank all the organizers of COSMO-2000 for an enjoyable
and stimulating conference, and for continuing the now traditional focus on
particle cosmology. In addition, I am indebted to the Korean Institute for Advanced Study (KIAS) 
for support during the conference and hospitality in Seoul.


\begin{thebibliography}{99}

\bibitem{Anton}
I.~Antoniadis, \PL {\bf B246}, 377 (1990);
J.~Lykken \PR {\bf D54} 3693 (1996).

\bibitem{ADD}
N.~Arkani-Hamed, S.~Dimopoulos and G.~Dvali, \PL {\bf B429},
263 (1998); I.~Antoniadis, {\em et al}, \PL {\bf B436},257 (1998).

\bibitem{HW}
P. Ho{\v r}ava and E. Witten, \NP {\bf B460}, 506 (1996);
\NP {\bf B475}, 94 (1996).

\bibitem{polchinski}
J.~Polchinski, \PRL {\bf 75},4724 (1995).

\bibitem{Landsberg}
G. Landsberg, ``Minireview on Extra Dimensions'', hep-ex/0009038 (2000).

\bibitem{collider}
I.~Antoniadis, K.~Benakli, M.~Quiros, \PL {\bf B331}, 313 (1994);
G.~Giudice, R.~Rattazzi and J.~Wells, \NP {\bf B544},3 (1999);
E.~Mirabelli, M.~Perelstein, M.~Peskin, \PRL {\bf 82}, 2236 (1999);
T.~Han, J.~Lykken, R.~Zhang, \PR {\bf D59} 105006 (1999).

\bibitem{KKconstraints}
S.~Cullen, M.~Perelstein, \PRL {\bf 83} (1999) 268.

\bibitem{our paper}
N.~Kaloper, J.~March-Russell, G.~Starkman and M.~Trodden, 
\PRL {\bf 85}, 928 (2000); [{\tt hep-ph/0002001}].

\bibitem{Thurston}
See for example:
W.P.~Thurston, {\em Three-Dimensional Geometry and
Topology} (Princeton UP, Princeton, 1997).

\bibitem{RS}
L.~Randall and R.~Sundrum, \PRL {\bf 83}, 3370 (1999); {\it ibid}
4690 (1999).

\bibitem{MostowPrasad}
G.~Mostow, Ann. Math. Stud.{\bf 78} (Princeton UP,
Princeton 1973); G.~Prasad, Invent. Math.{\bf 21} 255 (1973).

\bibitem{geomview}
{\tt http://thames.northnet.org/weeks/index/SnapPea.html}

\bibitem{snappea}
{\tt http://www.geomview.org/overview/}

\bibitem{Brooks}
R.~Brooks, private communication.

\bibitem{numeric}
N.~Cornish and  D.N.~Spergel [{\tt math.DG/9906017 }]
and references therein.

\bibitem{ADD3}
N.~Arkani-Hamed, S.~Dimopoulos and G.~Dvali,
\PR {\bf D59}, 086004 (1999).

\bibitem{ADKMR}
N.~Arkani-Hamed, S.~Dimopoulos, N.~Kaloper and J.~March-Russell, \NP {\bf B5667}, 189 (2000); 
[{\tt hep-ph/9903224}].

\end{thebibliography}
\end{document}